# Coupling of electronic and magnetic properties in $Fe_{1+y}(Te_{1-x}Se_x)$

J. Hu, T.J. Liu, B. Qian and Z.Q. Mao*

Department of Physics and Engineering Physics, Tulane University, New Orleans,

Louisiana 70118, USA

Abstract

We have studied the coupling of electronic and magnetic properties in $Fe_{1+y}(Te_{1-x}Se_x)$ via systematic specific heat, magnetoresistivity, and Hall coefficient measurements on two groups of samples with $y = 0.02$ and $0.1$. In the $y = 0.02$ series, we find that the $0.09 < x < 0.3$ composition region, where superconductivity is suppressed, has large Sommerfeld coefficient $\gamma$ (~55-65 mJ/mol $K^2$), positive Hall coefficient $R_H$ and negative magnetoresistance MR at low temperature, in sharp contrast with the $x$=0.4-0.5 region where $\gamma$ drops to ~ 26 mJ/mol $K^2$ and $R_H$/MR becomes negative/positive at low temperature. Dramatic changes of $\gamma$, as well as sign reversal in low-temperature $R_H$ and MR, are also observed across the $x$~0.1 boundary where the long-range antiferromagnetic order is suppressed. However, for the system with rich interstitial excess Fe ($y = 0.1$), where bulk superconductivity is suppressed even for $x$=0.4-0.5, the variations of $\gamma$, $R_H$ and MR with $x$ are distinct from those seen in $y = 0.02$ system: $\gamma$ is ~40 mJ/mol $K^2$ for $0.1 < x < 0.3$, and drops to ~ 34 mJ/mol $K^2$ for $x = 0.4$-$0.5$; $R_H$ and MR does not show any sign reversal as $x$ is increased above 0.3. We will show that all these results can be understood in light of the evolution of the incoherent magnetic scattering by $(\pi,0)$ magnetic fluctuations with Se concentration. In addition, with the suppression of magnetic scattering by magnetic field, we observed the surprising effect of a remarkable increase in the superconducting volume fraction under moderate magnetic fields for $x$=0.3-0.4 samples in the $y = 0.02$ system.

*zmao@tulane.edu

# I. INTRODUCTION

The interplay between magnetism and superconductivity in Fe-based superconductor systems is currently a subject of intensive studies. The iron chalcogenide $Fe_{1+y}(Te_{1-x}Se_x)$ is of particular interest due to its unique magnetic properties. While the parent compound $Fe_{1+y}Te$ shows antiferromagnetism with $(\pi,0)$ in-plane magnetic wave vector [1,2], the superconducting state obtained by Se substitution for Te displays spin resonance at $(\pi,\pi)$ [3,4]. This contrasts with iron pnictides in which both the parent compound's antiferromagnetism [5,6] and the doped samples' superconducting spin resonance [7-9] are characterized by the in-plane Fermi surface nesting wave vector $Q_n = (\pi,\pi)$. The evolution from $(\pi,0)$ magnetism to superconductivity with $(\pi,\pi)$ spin resonance in $Fe_{1.02}(Te_{1-x}Se_x)$ is associated with coexisting magnetic correlations at $(\pi,0)$ and $(\pi,\pi)$ [10]. Another remarkable difference between iron chalcogenide and iron pnictide superconductors is their phase diagrams. In iron pnictides, bulk superconductivity either emerges immediately following suppression of long-range $(\pi,\pi)$ antiferromagnetic (AFM) order [11,12], or coexists with it in a particular composition range [13-16]. In contrast, in $Fe_{1.02}(Te_{1-x}Se_x)$, bulk superconductivity does not appear immediately following the suppression of long-range $(\pi,0)$ AFM order. Instead, an intermediate phase with weak charge carrier localization appears between metallic AFM state and bulk superconducting phase [10].

In addition, another distinct characteristic of $Fe_{1+y}(Te_{1-x}Se_x)$ is that both the magnetism and superconductivity of this system are sensitively dependent on the Fe non-stoichiometry, which originates from the partial occupation of excess Fe at the interstitial sites of the Te/Se layer [1]. In the parent compound $Fe_{1+y}Te$, the long-range ($\pi$,0) AFM order can be tuned from commensurate to incommensurate when $y > 0.076$ [1]. As the long-range AFM order is suppressed by Se substitution for Te, interstitial Fe causes the magnetic Friedel-like oscillations and stabilizes the ($\pi$,0) short-range glassy magnetism [17]. Furthermore, rich interstitial Fe results in weak charge carrier localization [18] and consequently suppresses superconductivity even for $x$ =0.4-0.5 [18, 19].

In our previous work, ($\pi$,0) magnetic fluctuations are found to be antagonistic to superconductivity and associated with weak charge carrier localization in $0.1 < x < 0.3$ samples [10]. However, the microscopic origin of such effects has not been revealed. The clarification of this issue is crucial for the understanding of the unusual phase diagram of $Fe_{1+y}(Te_{1-x}Se_x)$ system, and may shed light on the mechanism responsible for the perplexing interplay between the magnetism and superconductivity in Fe-based superconductors. In this article, we have investigated the dependence of electronic properties on ($\pi$,0) magnetic fluctuations in $Fe_{1+y}(Te_{1-x}Se_x)$. Our central goal is to address the role of the ($\pi$,0) magnetic fluctuations in transport mechanism. We have measured superconducting volume fraction $V_{sc}$, Sommerfeld coefficient $\gamma$, Hall coefficient $R_H$ and magnetoresistivity (MR) as a function of Se concentration for two groups of single crystal samples with interstitial excess Fe content $y \sim 0.02$ and 0.10. In samples with less interstitial Fe ($y \sim 0.02$), we found that all these quantities exhibit unusual evolutions with Se content. In the superconducting region, $V_{SC}$, as well as the normal state metallicity, strongly couples with the variation of $\gamma$ and $R_H$; both $V_{SC}$ and the normal state metallicity increase significantly as $\gamma$ and $R_H$ sharply drop near $x$=0.3-0.4 where MR also exhibits a sign reversal. Furthermore, we observed the surprising increase of $V_{SC}$ by moderate magnetic fields for $x=0.3$-$0.4$ samples. These results reveal the mechanism for the superconductivity suppression in the $0.1<x<0.3$ region observed in our previous work [10]: $(\pi,0)$ magnetic fluctuations cause incoherent magnetic scattering which reduces the charge carriers' mobility and results in charge carrier localization. This mechanism is further corroborated by the results obtained on the samples with rich excess Fe ($y \sim 0.10$), where we observed much smaller variations in $\gamma$, $R_H$ and MR as compared to the $y = 0.02$ system.

## II. EXPERIMENT

Two groups of $Fe_{1+y}(Te_{1-x}Se_x)$ single crystal samples with nominal interstitial Fe content $y = 0$ and $y = 0.14$ were synthesized using a flux method as reported elsewhere [9]. The actual content of interstitial Fe measured by energy-dispersive X-ray spectrometer (EDXS) is $y \sim 0.02$ and $0.1$, respectively, for these two groups of samples. As indicated above, since Fe non-stoichiometry greatly affects the magnetic and electric properties of this system [1, 18, 19], each sample selected for measurements was carefully screened by EDXS. The Fe and Se concentrations shown in the following figures and text are all measured values by EDXS. We measured resistivity with a four-probe method, Hall effect with a five-probe method and specific heat with an adiabatic relaxation technique using a physical property measurement system (Quantum Design).

## III. RESULTS AND DISCUSSION

**A. Studies on samples with less interstitial Fe: $Fe_{1.02}(Te_{1-x}Se_x)$**

## A.1. Specific heat of Fe$_{1.02}$(Te$_{1-x}$Se$_x$)

We present our results for samples with less interstitial excess Fe ($y \sim 0.02$) in this section. Figure 1(a) shows the phase diagram for Fe$_{1.02}$(Te$_{1-x}$Se$_x$) (denoted as Fe$_{1.02}$ below) determined by specific heat measurements. The superconducting transition temperatures $T_c$ and the Néel temperatures $T_N$ probed by anomaly peaks of specific heat (see Fig. 2(a) and 2(b)) are exactly consistent with our phase diagram previously established by resistivity, magnetization, and neutron scattering measurements [9]. While the AFM order is fully suppressed near $x \sim 0.1$, a superconducting anomaly peak in specific heat is not observed until $x$ is increased above 0.3. The gradual decrease of the superconducting transition width with the increase of $x$ seen in Fig. 2(b) indicates that a homogeneous superconducting transition does not emerge until $x$ reaches 0.4-0.5 (see below for detailed discussions). The specific heat data allows for precise evaluation of superconducting volume fraction $V_{SC}$, i.e. $V_{SC} = (\gamma - \gamma_{res})/\gamma$, with $\gamma$ and $\gamma_{res}$ being the Sommerfeld coefficient and residual electronic specific heat coefficient respectively. $\gamma_{res}$ can be obtained by fitting the specific heat data at temperatures well below $T_c$ to $C/T = \gamma_{res} + \beta T^2$ as shown in Fig. 3a, while $\gamma$ can be derived via separating electronic specific heat from phonon contribution using the specific heat of a Cu-doped, non-superconducting Fe$_{1.02}$(Te$_{1-x}$Se$_x$) sample as a reference (see ref. 20 for detailed analyses). We examined the evolution of $V_{SC}$ with Se content $x$ following the analyses reported in ref. 20. As summarized by the contour plot in Fig. 1(a), $V_{SC}$ estimated from specific heat shows a systematic increase for $x > 0.3$ (Region III). $V_{SC}$ is ~8% near $x = 0.3$ and rises rapidly for $x > 0.3$, up to ~90% for $x = 0.4$-0.5 samples. The $V_{SC}$ previously evaluated from diamagnetism for $0.29 < x < 0.4$ in ref. 10 was overestimated due to the magnetic shielding effect which usually occurs for inhomogeneous superconductors [20, 21], which will be discussed in greater details below. For $x < 0.09$ (Fig. 2(a)) and $0.09 < x < 0.29$ (the specific heat data of which were reported in our earlier work [10]), the specific heat data does not exhibit any remarkable signatures associated with bulk superconductivity.

In Fig. 1(a) we have also included the contour plot of the derivative of in-plane resistivity $d\rho_{ab}/dT$ for Region II and III, as well as the contour plot of the exponent $n$ of the temperature-dependent term of $\rho_{ab}$ ($\propto T^n$) for the AFM state in Region I. Since the superconducting transition probed in resistivity in Regions I and II only represents a trace of superconductivity caused by chemical inhomogeneity (see below for more detailed discussions), we do not include it in the phase diagram of Fig. 1(a), and the data of $d\rho_{ab}/dT$ and $n$ below $T_c$ are both extrapolated from the data immediately above $T_c$. The weakly localized state represented by a negative $d\rho_{ab}/dT$ in Region II (Green) sharply contrasts with the metallic state in Region I (Red) and III (Yellow). The weakly localized-to-metallic crossover boundary between Region II and III occurs simultaneously with the remarkable increase of $V_{SC}$ near $x \sim 0.35$, as seen in the $V_{SC}$ contour map in Fig. 1(a). This indicates that homogeneous superconducting phase develops only when significant charge carrier delocalization occurs.

Since the superconductivity couples with the normal state properties as revealed by the phase diagram in Fig. 1(a), we further studied the normal state properties via examining the variation of the normal state Sommerfeld coefficient $\gamma$ with $x$ from specific heat measurements (see Fig. 1(b)). For superconducting samples in Region III, as noted above, we derived $\gamma$ through separating electronic specific heat from phonon contribution using non-superconducting $(Fe_{0.9}Cu_{0.1})(Te_{1-x}Se_x)$ ($x$ =0.37 and 0.43) samples as references [22]. For those samples in Regions I and II, $\gamma$ is quoted from our previous work [10]. As seen in Fig. 1(b), $\gamma$ first rises significantly with the suppression of AFM order in Region I, then shows a broad maximum over Region II (~ 55-65 mJ/mol K$^2$) , and finally decreases dramatically near $x$ ~ 0.35 where both the superconductivity and the normal state metallicity are markedly enhanced. $V_{SC}$ does not reach a maximum until $\gamma$ drops to a minimal value of ~26 mJ/mol K$^2$ near $x$ = 0.4-0.5. Such an evolution of $\gamma$ is consistent with a previous report [23] and distinct from $Ba(Fe_{1-x}Co_x)_2As_2$ where $\gamma$ maximizes near optimal doping [24].

From specific heat measurements under magnetic fields as shown in Fig. 3(a), we have observed surprising, unusual field dependence of electronic specific heat for superconducting samples in Region III. The field-induced change in the electronic specific heat $\Delta\gamma/\gamma$ ($\Delta\gamma=\gamma(B)-\gamma(0)$) at the zero temperature limit decreases significantly with an increase of magnetic field after reaching a maximum (Fig. 3(b)). In general, $\Delta\gamma$ increases with magnetic field in different manners for different paring symmetries owing to the orbital depairing effect. Our observation of the unusual field dependence of $\Delta\gamma$ clearly does not fit into this expected scenario, but can be attributed to field-induced superconductivity enhancement as discussed below.

## A.2. Hall effect and Hall angle of $Fe_{1.02}(Te_{1-x}Se_x)$

In addition to specific heat, we also preformed systematic Hall effect measurements to study the normal state properties. The Hall coefficient $R_H$ of $Fe_{1.02}$ samples exhibits distinct temperature dependences within different composition regions, as shown in Fig. 4. For Region I, our earlier work showed that $R_H$ sharply drops from a positive to negative value across $T_N$ [10]. In contrast, in Region II, $R_H$ remains positive and increases monotonically with decreasing temperature (Fig. 4(a)). Such low temperature enhancement of $R_H$ is gradually suppressed with a further increase of Se content in Region III, and $R_H$ returns to negative values at low temperatures as $x$ is increased above 0.4 (Fig. 4(b)). We have presented the dependence of $R_H(20\ K)/R_H(300\ K)$ on $x$ for $Fe_{1.02}$ samples in Fig. 1(c), which shows a broad maximum over Region II and sign reversals across both phase boundaries. Such variation of $R_H$ with Se concentration is consistent with the results obtained previously on polycrystalline samples [25].

Sharp increases of $R_H$ and $\gamma$ near the AFM phase boundary ($x \sim 0.1$, Fig. 1(b) and (c)) should be associated with the massive band reconstruction caused by the simultaneous structural and AFM transition, as evidenced by photoemission measurements [26]. However, drastic decreases in $R_H$ and $\gamma$ near $x = 0.35$ should neither be purely structurally driven, nor caused by charge carrier doping due to the absence of structural transitions near $x \sim 0.35$ [27] and the isovalent substitution; they cannot be attributed to the variation of crystallographic disorder scattering either, since the charge carrier localization occurs only in the $0.1 < x < 0.3$ region, which is less disordered. Then what is the origin for these unusual variations of $\gamma$ and $R_H$? Because the Fermi surface of Fe chalcogenide consists of both electron and hole pockets [28, 29] with strong correlation being possibly involved[22, 30, 31], we cannot directly correlate $\gamma$ and $R_H$ with the density of states at Fermi level (DOS($E_F$)) and the charge carrier density $n$ as in a non-interacting single-band system where $\gamma \propto$ DOS($E_F$) and $R_H = -1/(ne)$. For a multiple-band system, $R_H$ is not only dependent on carrier density, but is also associated with charge carriers' mobility. Multiple-band signature of Fe$_{1+y}$(Te$_{1-x}$Se$_x$) has indeed been probed in thermopower measurements [25, 32].

To examine the variation of charge carrier mobility with Se doping, we further investigate the scattering rate through the analyses of Hall angle $\theta_H = \cot^{-1}(\rho_{xx}/(\rho_{xy}/B))$ of three typical samples in different phase regions. As shown in Fig. 5(a)-(c), $\cot\theta_H$ follows the same quadratic temperature dependence as the longitudinal resistivity $\rho_{xx}$ does below $T = 25$ K for Fe$_{1.02}$Te (Fig. 5(a)), but exhibits linear $T$-dependence in the $x = 0.20$ sample whose $\rho_{xx}$ shows a remarkable upturn with decreasing temperature (Fig. 5(b)). For the $x = 0.45$ sample, though $\rho_{xx}$ and $\cot\theta_H$ still follow different $T$-dependences, they both decrease monotonically with temperature (Fig. 5(c)). Since $\rho_{xx}$ and $\cot\theta_H$ probe longitudinal and transverse scattering rates, respectively, the evolutions of $\rho_{xx}$ and $\cot\theta_H$ described above suggests that the AFM state in Region I has a single relaxation rate for charge carrier motion, while in Region II an additional scattering takes place, leading to the distinct scaling for $\rho_{xx}$ and $\cot\theta_H$. Such additional scattering weakens in Region III, where the disparity between the temperature dependence for $\rho_{xx}$ and $\cot\theta_H$ decreases.

What is the origin of such additional scattering? Given that $Fe_{1+y}(Te_{1-x}Se_x)$ is characterized by a coexistence of magnetic correlations at $(\pi,0)$ and $(\pi,\pi)$, and that the superconductivity suppression is accompanied by charge carrier localization in the $0.1 < x < 0.3$ region where the $(\pi,0)$ magnetic correlation is strong [10], the magnetic scattering by the $(\pi,0)$ magnetic correlation appears to be a possible mechanism. In fact, all of our experimental data and the analyses presented below point to this scenario. Furthermore, since the $(\pi,0)$ magnetic fluctuations show diffusive characteristics [1, 33], the magnetic scattering by $(\pi,0)$ magnetic fluctuations is expected to have an incoherent component. The fact that the decrease of the disparity between the temperature dependences of $\rho_{xx}$ and $\cot\theta_H$ takes place when the $(\pi,0)$ magnetic fluctuations are suppressed from Region II to III agrees well with the magnetic scattering mechanism. Distinct temperature dependences between $\rho_{xx}$ and $\cot\theta_H$ resulting from magnetic scattering is indeed a common phenomenon, often observed in a system in close proximity to a magnetic instability, such as high-$T_c$ cuprates [34] and $V_{2-y}O_3$ [35].

### A.3 Magnetoresistance of $Fe_{1.02}(Te_{1-x}Se_x)$

To seek further evidence for the magnetic scattering, we preformed systematic MR measurements. As shown in Fig. 6, the $0.1 < x < 0.4$ samples show negative MR at low temperatures, in contrast with the positive MR of $Fe_{1.02}Te$ and the $x \geq 0.4$ samples. Although Kondo origin of negative MR could not be ruled out and requires further study, we found that the negative MR is essentially dependent on the $(\pi,0)$ magnetic fluctuations in this system. First, as seen in the inset in Fig. 6(a), the MR decreases in magnitude with increasing Se content for $0.1 < x < 0.4$, which is consistent with the trend that the $(\pi,0)$ magnetic fluctuations become suppressed with an increase of $x$ [10]. Secondly, the striking negative MR for $0.1 < x < 0.2$ samples is observed only below the temperature where the $(\pi,0)$ fluctuations develop. For example, the striking increase of the MR magnitude of the $x = 0.15$ sample occurs below 40K (see the inset in Fig. 6(b)) and the $(\pi,0)$ short-range magnetic order was found to take place below almost the same temperature [36]. Such correlations between the magnetotransport properties and the $(\pi,0)$ magnetism implies the magnetic origin of the negative MR of the underdoped samples.

The possible origin of such negative MR in the $0.1 < x < 0.4$ samples can be attributed to the suppression of the incoherent magnetic scattering by magnetic field. The AFM order at $(\pi,0)$ in the parent compound $Fe_{1+y}Te$ originates from local moments; this has been demonstrated by growing experimental evidence, such as the Curie-Weiss behavior of magnetic susceptibility [37], the large ordered magnetic moment ($\sim 2.1 \mu_B$/Fe) [10] and the magnetic excitation spectra described by the Heisenberg model [38]. Thus, the short-range $(\pi,0)$ AFM order in the samples with partial Te being replaced by Se is naturally expected to have similar local moment origin. On the other hand, since the AFM ordered state of $Fe_{1+y}Te$ exhibits metallic transport properties [18, 37], the coexistence of itinerant charge carriers and local moment magnetic order is an effective model to understand the AFM, metallic state of $Fe_{1+y}Te$. In fact, the neutron scattering measurements by Zaliznyak *et al.* [39] revealed remarkable evidence for the interaction between itinerant charge carriers and local moments. Such interaction is naturally expected to exist in those $0.1 < x < 0.4$ samples and should be responsible for the additional scattering reflected in Hall angle analyses. Magnetic field suppresses this additional magnetic scattering, leading to negative MR as discussed below.

**A.4. Possible origin of negative magnetoresistance and superconductivity enhancement under magnetic fields**

The ($\pi$,0) spin fluctuations in $x > 0.1$ samples freeze into a spin glassy state at low temperatures [36]. Such ($\pi$,0) glassy magnetism gradually weakens with an increase of Se content and the weak glassy magnetism survives even near $x = 0.4\text{-}0.5$. Recent neutron scattering measurements by Thampy *et al*.[17] revealed that the ($\pi$,0) glassy magnetism arises from the magnetic Friedel-like oscillations surrounding interstitial Fe. That is, a spin cluster involving more than 50 neighboring Fe ions on the Fe plane nucleates around interstitial Fe. Since the magnetic moments of Fe ions inside the cluster follow Friedel-like oscillations and the magnitude of moment decays with increasing distance from the interstitial Fe, the moments cannot fully cancel out via AFM coupling, thus resulting in residual magnetization in the spin cluster. Such spin cluster structure seeds and stabilizes the ($\pi$,0) magnetism in $Fe_{1.02}(Te_{1-x}Se_x)$.

Given that the charge carriers from itinerant bands interact with local moments participating in the ($\pi$,0) short-range order, the negative magnetoresistance may be attributed to the weak polarization of these spin clusters under magnetic field. Specifically, the magnetization of spin clusters can have an essential effect on the spin state of itinerant carriers; thus the incoherent spin scattering is likely involved in the transport between spin clusters. The applied magnetic field may suppress such random spin scattering via polarizing spin clusters, thus reducing resistivity.

With the understanding of the suppression of magnetic scattering by magnetic field, the significant decrease of electronic specific heat induced by the magnetic field shown in Fig. 3 can reasonably be attributed to an enhancement of superconductivity; this is further supported by the systematic variation of $\Delta\gamma$ with $x$ (see Fig. 3(b)). The initial increase of $\Delta\gamma$ at low field can be ascribed to the field-induced depairing effect. The superconductivity enhancement caused by field is more significant for $0.3 < x < 0.4$ samples. For instance, in the $x = 0.33$ sample, $V_{SC}$ rises from ~25% for zero field up to ~50% for 9 T, in sharp contrast with the x=0.4-0.5 samples where $(\pi,0)$ fluctuations are already greatly suppressed by Se substitution for Te, and thus, the field-induced superconductivity enhancement is less noticeable (see the inset to Fig. 3(b)).

**A.5. Possible mechanism for the unusual variations of $R_H$ and $\gamma$ with Se concentration**

In light of the incoherent magnetic scattering proposed above, the variation of $R_H$ in Fig. 1(c) can be understood well. In general, we can expect that incoherent magnetic scattering reduces charge carrier mobility. Given that the $(\pi,0)$ magnetic correlation is much stronger in the $x = 0.1$-0.3 samples than in $x = 0.4$-0.5 samples, as stated above, the $x = 0.1$-0.3 samples should have much lower charge carrier mobility than the x = 0.4-0.5 samples; this has actually been reflected in the contour plot of $d\rho_{ab}/dT$ shown in Fig. 1(a), which shows negative values for the $0.1 < x < 0.3$ region, but positive values for the $x$ = 0.4-0.5 region. The decrease of $R_H$ from positive values in the 0.1< $x$ <0.3 region to negative values in the $x$>0.4 region (see Fig. 1(b)) implies that electrons gain greater mobility when the (π,0) correlation becomes weak. This agrees with recent photoemission spectroscopy measurements which indeed reveal that the itinerancy of the electron bands near $M$, as well as their spectral weight, increase significantly as $x$ is increased above 0.4 [40]. Lower charge carrier mobility accounts for the superconductivity suppression and the weakly localized state over the 0.1< $x$ < 0.3 region (*i.e.* Region II in Fig. 1(a)).

Since the Sommerfeld coefficient $\gamma$ is a measure of the effective mass of quasi-particles, which is associated with the energy band curvature, the evolution of $\gamma$ with Se concentration shown in Fig. 1(b) may provide insights into the dependence of band structure on Se concentration. Recent photoemission measurements show that the Fe 3*d* orbital involves significant band renormalization in samples with 0.1< $x$ <0.35, and the evolution of derived effective mass with Se concentration is in good agreement with the variation of $\gamma$ seen in our experiments [40]. As described above, the evolution of $\gamma$ with Se concentration couples systematically with the evolutions of other quantities including $R_H$, MR, normal state metallicity and superconducting volume fraction $V_{SC}$. The bulk superconducting state with the maximum $V_{SC}$ is obtained only when $\gamma$ reaches a minimum value in the $x$ = 0.4-0.5 samples where low-temperature $R_H$ is negative and MR is positive. Given that $R_H$, MR, normal state metallicity and $V_{SC}$ are all associated with the incoherent scattering caused by (π,0) magnetic fluctuations, the enhanced band renormalization in the 0.1< $x$ <0.35 samples may also be associated with the magnetic scattering. Although further theoretical and experimental studies are needed to clarify the underlying mechanism of such band renormalization, its relevance with the magnetic scattering is also revealed by our studies on samples with rich interstitial Fe as will be shown later.

**A.6. Transport mechanism of the parent compound $Fe_{1.02}Te$**

From the role of $(\pi,0)$ magnetic correlation in transport mechanism discussed above, the electronic transport properties of the parent compound $Fe_{1.02}Te$ can be understood well. In general, magnetic fluctuations should weaken abruptly when they develop into a static long-range order. This should apply to $Fe_{1.02}Te$ for its AFM transition. At temperature above $T_N \sim$ (72 K), the system should possess strong $(\pi,0)$ magnetic fluctuations, which should results in striking incoherent magnetic scattering according to the discussions given above. Since incoherent magnetic scattering leads to charge carrier localization, the non-metallic temperature dependence of resistivity seen in $Fe_{1+y}Te$ above $T_N$ (see Fig. 4 in ref. 18) can naturally be understood. When the long-range $(\pi,0)$ AFM order forms below $T_N$, the magnetic scattering mostly becomes *coherent* due to weakened fluctuations, thus yielding coherent electron wave. This may explain the non-metal-to-metal transition across $T_N$, as well as the Fermi liquid behavior in the ground state [18]. The Lorentz-type orbital magnetoresistance ($\Delta\rho \propto B^2$) seen in $Fe_{1.02}Te$ (see Fig.6a), as well as the same quadratic temperature dependence of Hall angle and longitudinal resistivity (see Fig. 5a), also indicate that incoherent magnetic scattering is nearly negligible in the electronic transport process at low temperature.

**A.7. Origin of the weak superconductivity in the 0.1< $x$ <0.3 region**

We have noticed that our phase diagram shown in Fig. 1(a) looks different from those reported by several groups [41-43]. In our phase diagram, the bulk superconducting phase is separated from the long-range AFM ordered phase by a weakly localized phase within the 0.1< $x$ <0.3 range. The weakly localized region is also accompanied by a trace of superconductivity [10]. However, the phase diagrams reported in refs. [41-43] highlight coexistence of superconductivity and antiferromagnetism. Such inconsistence can easily be reconciled by considering chemical inhomogeneity, which is unavoidable for any alloy systems such as $Fe_{1+y}(Te_{1-x}Se_x)$. In fact, Hu *et al*. [44] have observed remarkable chemical inhomogeneity in $Fe_{1+y}(Te_{1-x}Se_x)$ in scanning transmission electron microscopy (STEM) and electron energy-loss spectroscopy (EELS) analyses. They found that Te concentration shows ~20% (or low) fluctuations from the average local composition, which leads to nanoscale phase separation of compositions. Given that neutron scattering measurements have established that Se/Te controls the strength of antiferromagnetic correlation at $(\pi,0)$ and the increase of Se concentration weakens $(\pi,0)$ magnetic correlation [10, 43], compositional phase separation must lead to an inhomogeneous magnetic state (*i.e.* magnetic phase separation): the local areas with richer Se should have weaker $(\pi,0)$ magnetic correlation than those local areas with richer Te. This is evidenced by the observation of magnetic glassy phase [36, 45]. Given that the $(\pi,0)$ magnetic correlation is antagonistic to superconductivity as discussed above, the inhomogeneous magnetic state with $(\pi,0)$ magnetic correlation should result in inhomogeneous superconductivity; this is evidenced in our specific heat measurements which show that the superconductivity in the $0.1< x <0.3$ range is just a trace, with very low superconducting volume fraction (i.e. $V_{SC}$ <1-3%) as shown above [10].

In samples containing inhomogeneous superconductivity with a non-superconducting phase being involved, superconducting responses probed by different experimental techniques could be very different. When superconductivity exists as a minor phase in a given sample, the non-bulk probe for superconductivity, such as resistivity measurements, can still detect superconductivity, while the bulk-sensitive measurement such as specific heat may not show any noticeable signature of superconductivity. This is exactly what have observed in the $Fe_{1.02}(Te_{1-x}Se_x)$ samples with $0.1< x < 0.3$. In these samples, we observed zero-resistance superconducting states in resistivity measurements, but did not find any superconducting anomalies in specific heat measurements (see the resistivity and specific heat data in the Fig. 3 of ref. 10). Such inconsistence can be easily understood. As long as the minor superconducting phase forms a continuously connected network, the sample could show a zero-resistance superconducting state though its $V_{SC}$ is only a few percent. Ac susceptibility, which is often used to characterize superconducting properties, is also a non-bulk probe. What is probed in ac susceptibility is actually a screening effect. A well connected network of minor superconducting phase could lead to strong screening current in spite of low $V_{SC}$, which could in turn result in strong diamagnetic response. In this case, diamagnetic response does not precisely reflect the actual $V_{SC}$. Therefore, ac susceptibility is normally not used to evaluate $V_{SC}$. Instead, the dc susceptibility measured at very low field and with zero-field cooling history is commonly used to estimate $V_{SC}$ (= $-4\pi\chi$), which is based on the perfect diamagnetism of superconducting state. However, when a non-superconducting phase or voids are included in a given sample (this is the case for $Fe_{1+y}(Te_{1-x}Se_x)$ with $0.1 < x < 0.3$), $V_{SC}$ could easily be overestimated since the superconducting phase may shield non-superconducting phases/voids. This has been clearly addressed by Ando & Akita [21]. Here we give an example which was reported in our previous work [20] where we found that an oxygen-annealed $Fe_{1.02}Te_{0.8}Se_{0.2}$ sample displays the strongest diamagnetism with $-4\pi\chi \sim 1$, but has only $V_{SC} = 16\%$ (see Fig. 1(a) and 4(b) in ref. 20), indicating that the shielding effect could cause a significant overestimate of $V_{SC}$ for a sample containing a non-superconducting phase. We notice that the superconductivity shown in the phase diagrams reported in refs. [41-43] are determined by ac/dc susceptibility. Therefore, it is not surprising to observe signatures of superconductivity in the $0.1 < x < 0.3$ composition region. However, the actual $V_{SC}$ is very low (<1-3%) for this composition region as shown in our specific heat measurements (see above), which should be attributed to the phase separation arising from the compositional inhomogeneity as discussed above.

For an inhomogeneous superconducting system like $Fe_{1+y}(Te_{1-x}Se_x)$, the most precise estimation of $V_{SC}$ can be made through specific heat measurements as indicated above, since it probes thermodynamic properties of superconducting state, which is bulk sensitive. In this approach, $V_{SC}$ is estimated from the ratio of the residual electronic specific heat $\gamma_{res}$ of superconducting state to the normal state Sommerfeld coefficient $\gamma$ as indicated above. $\gamma_{res}$ should be zero for a given superconductor with a 100% volume fraction and has a finite value for a superconductor containing a non-superconducting phase due to the existence of unpaired electrons. As shown in Fig. 3a, $\gamma_{res}$ is rather small (~1.8 mJ/mol K$^2$) for the $x = 0.43$ sample which has $V_{SC} = 92\%$, but increases to ~26.9 mJ/mol K$^2$ as $x$ is decreased to 0.33, indicating dramatic decrease of $V_{SC}$. As indicated above, specific heat does show any feature associated with superconductivity for $0.1 < x < 0.3$ (see Fig. 3 in ref. 10). The superconductivity probed in resistivity or susceptibility in this range can only be attributed to phase separation due to chemical inhomogeneity as stated above. From Fig. 1(a), it can be seen that $V_{SC}$ rises to 8% at $x$~0.3, and steeply increases up to >90% for $x \geq 0.4$; this indicates that the superconductivity is still considerably inhomogeneous for $0.3 < x < 0.4$ despite relatively high $V_{SC}$, but becomes almost homogeneous for $x = 0.4$-0.5; this is further evidenced by the observation that the superconducting transition width probed in specific heat (see Fig.2(b)) is considerably broad for $0.3 < x < 0.4$ but narrow for $x > 0.4$.

**A.8. Effect of interstitial Fe on the phase diagram of Fe$_{1+y}$(Te$_{1-x}$Se$_x$)**

The superconductivity of Fe$_{1+y}$(Te$_{1-x}$Se$_x$) is not only tuned by the Se/Te ratio, but also controlled by interstitial Fe content $y$. Rich interstitial Fe has been shown to suppress superconductivity even for $x = 0.4$-0.5 samples [18] and Fe$_{1+y}$Se [19], the origin of which is that it enhances the ($\pi$,0) magnetic correlation [43], which is destructive to superconductivity. This is further evidenced by recent neutron scattering measurements that revealed that the ($\pi$,0) short-range order nucleates around interstitial Fe [17]. Since the superconductivity of Fe$_{1+y}$(Te$_{1-x}$Se$_x$) depends on interstitial Fe, the phase diagram of this system can be altered by controlling the interstitial excess Fe content $y$ [20]. The phase diagram we show in Fig. 1(a) are based on as-grown single crystals with ~2% interstitial

Fe. Interstitial Fe can be reduced by annealing the samples in various oxidation agents such as $O_2$, $I_2$, $N_2$ and air [20, 46-48]. Thus the phase diagram established using annealed samples shows enhanced superconductivity for $0.1 < x < 0.3$. Our previous specific heat measurements show that $V_{SC}$ can increase to 10-30% in the $0.1 < x < 0.3$ range for $N_2$-annealed samples [20]. We also note that several groups argued that $Fe_{1+y}(Te_{1-x}Se_x)$ samples annealed in air and $O_2$ have coexistence of superconductivity and magnetism in the $x < 0.3$ range [48, 49]. However, superconductivity characterization in those papers is based on magnetic susceptibility and resistivity measurements, which are not bulk sensitive for superconductivity as stated above. While annealing can lead to partial interstitial Fe deintercalation and enhance superconductivity to some extent, the superconductivity still involves significant phase separation for the $x < 0.3$ range, as we have demonstrated in our $N_2$ and $O_2$ annealed samples [20]. In addition to annealing, interstitial Fe can also be reduced by growing single crystals using nominal composition with deficient Fe. For example, Viennois [50] and Bendele *et al.* [43] observed enhanced superconductivity in $x = 0.2-0.3$ samples synthesized with nominal $y= -0.1$ and $-0.05$. On the other hand, the entire sample series does not show bulk superconductivity when $y$ is increased above 0.1 as shown below.

Another important characteristic associated with interstitial Fe is that its content may depend on the Se/Te ratio [51, 52]. The Te-rich samples contain more excess Fe. This may be associated with the fact that Se and Te have distinctly different heights from the Fe plane as revealed by Tegel *et al.* [53]. The large height of Te may require more interstitial Fe to stabilize the structure. The first principle calculations by Moon *et al.* [54]

show that the anion height in $Fe_{1+y}(Te_{1-x}Se_x)$ is a key factor in determining the nature of magnetic correlation. The larger anion height favors the $(\pi,0)$ ordering, while the smaller anion height leads to $(\pi,\pi)$ ordering. This result provides microscopic interpretation for the observation that the $(\pi,0)$ is tuned weak by Se substitution for Te [10] and the $(\pi,\pi)$ spin resonance develops in the superconducting state for $x = 0.4$-$0.5$ samples [3,4]. The samples used for establishing our phase diagram is synthesized with the $Fe_{1.0}(Te_{1-x}Se_x)$ nominal composition. Since we have been aware that interstitial Fe suppresses superconductivity [18], we intentionally chose the as-grown samples with minimum interstitial Fe for our experiments through EDXS measurements. All samples used for establishing the phase diagram in Fig. 1a have interstitial Fe $y \sim 0.02$. Given that EDXS has a limited resolution, our measurements cannot tell if $y$ varies with Se concentration though this is most likely the case.

**B. Studies on samples with rich interstitial Fe: $Fe_{1.1}(Te_{1-x}Se_x)$**

**B.1. Resistivity and specific heat of $Fe_{1.1}(Te_{1-x}Se_x)$**

Given that the $(\pi,0)$ magnetic correlation plays a critical role in suppressing superconductivity for $0.1 < x < 0.3$ and can be enhanced by increasing interstitial Fe, the argument of the magnetic scattering caused by $(\pi,0)$ fluctuations can be further examined using samples with rich interstitial Fe. In our previous work, we have investigated the effect of interstitial Fe on superconductivity using a sample with $x = 0.36$ [18]. Here we expand this study to the samples covering the entire phase diagram. We have prepared another group of sample with rich excess Fe, *i.e.*, $Fe_{1.1}(Te_{1-x}Se_x)$ (denoted as $Fe_{1.1}$ below ) for this purpose and conducted systematic resistivity, specific heat, Hall effect, and MR measurements. Our intention is to clarify how the increase of interstitial Fe affects the evolutions of $\gamma$, $R_H$ and MR with Se concentration and compare them with those results obtained in the Fe$_{1.02}$ samples. This would help us better understand how interstitial Fe and ($\pi$,0) magnetic fluctuations affect the electronic structure and transport mechanism.

Figure 7 presents in-plane resistivity $\rho_{ab}$ for Fe$_{1.1}$ samples. All samples show nonmetallic behavior at low temperatures, in sharp contrast with the Fe$_{1.02}$ system where only the 0.1< $x$ < 0.35 region shows non-metallic behavior. This indicates that the increase of interstitial Fe tends to localize charge carriers even for the $x$ = 0.4-0.5 region, consistent with our previous observation in the Fe$_{1.11}$(Te$_{0.64}$Se$_{0.36}$) sample [18]. This could be explained in terms of the magnetic scattering by the ($\pi$,0) magnetic fluctuations discussed above. We have pointed out in section A that in the system with less interstitial Fe, the ($\pi$,0) magnetic correlation is significantly weakened when Se concentration is increased above $x$ = 0.3. However, in the system with rich interstitial Fe, the ($\pi$,0) magnetic correlation was found to be enhanced even in the $x$ > 0.30 region [10]. In other words, the ($\pi$,0) magnetic fluctuations do not weaken strikingly with the increase of Se concentration in the Fe$_{1.1}$ system as it does in Fe$_{1.02}$ system. Therefore the magnetic scattering by the ($\pi$,0) magnetic fluctuations remains strong even in the $x$ > 0.4 range for the Fe$_{1.1}$ system. For the AFM ordered state in the $x$ < 0.1 region, since interstitial Fe carries magnetic moments and participates in the ($\pi$,0) order [17], rich interstitial Fe acts as perturbations and should increase the incoherence of magnetic scattering, which results in reduction of coherence of electronic states. This accounts for the non-metallic behavior induced by interstitial Fe in the AFM state (see the data for $x = 0$ and 0.05 in Fig. 7(a)).

Although all samples containing Se in the $Fe_{1.1}$ series show superconducting transitions in their resistivity data, specific heat measurements on them do not show any signature of bulk superconductivity (see Fig. 8), consistent with our previously reported result obtained on $Fe_{1.11}(Te_{0.64}Se_{0.36})$ [18]. Their low temperature specific heat data can be approximately fitted to $C/T = \gamma + \beta T^2$. We present two examples of fitting in the inset of Fig. 8(a). The samples with high Se concentrations exhibit slight deviation from fitting below 4.5 K owing to the presence of non-bulk superconductivity. The Sommerfeld coefficients $\gamma$ obtained from these fittings has been added to Fig. 1(b) for the comparison with those of $Fe_{1.02}$ samples. As seen from Fig. 1(b), the increase of interstitial Fe leads to a drastic effect on $\gamma$. The variation of $\gamma$ with Se concentration in the $Fe_{1.1}$ series is much less than that in the $Fe_{1.02}$ series. As indicated above, the $Fe_{1.02}$ system is characterized by large $\gamma$ values (~ 55-65 mJ/mol K$^2$) in the $0.1 < x < 0.3$ region but relatively small $\gamma$ (~ 26 mJ/mol K$^2$) in the $x > 0.4$ region. However, for the $Fe_{1.1}$ system, although $\gamma$ also shows a small drop when Se concentration $x$ is increased above 0.35, the $\gamma$ value remains around 34 mJ/mol K$^2$ even for $x > 0.4$, about 30% greater than those of $x$ =0.4-0.5 samples in the $Fe_{1.02}$ series. On the contrary, the $0.1 < x < 0.3$ composition region has much smaller $\gamma$ values (~ 40 mJ/mol K$^2$) than the same composition region in the $Fe_{1.02}$ series.

Our above discussions have suggested that the enhanced band renormalization in the $0 < x < 0.3$ region of the $Fe_{1.02}$ series is associated with the magnetic scattering. The evolution of $\gamma$ of the $Fe_{1.1}$ series provides further support for this argument. Since the $(\pi,0)$ short-range magnetic order nucleates around interstitial Fe [17] and interstitial Fe increases the magnetic correlation length [43], the increase of interstitial Fe would lead to enhanced $(\pi,0)$ magnetic correlation so that the $(\pi,0)$ magnetism does not weaken significantly with Se substitution for Te even for $x > 0.3$ in the $Fe_{1.1}$ system as it does in the $Fe_{1.02}$ system. This may account for the less significant drop in $\gamma$ near $x \sim 0.35$ for $Fe_{1.1}$ system as shown in Fig. 1(b). The slight increase of $\gamma$ across the $x \sim 0.1$ boundary in the $Fe_{1.1}$ system, which is in stark contrast with the sharp peak of $\gamma$ near $x \sim 0.1$ in the $Fe_{1.02}$ system, may imply that when $x$ is increased above 0.1, the $(\pi,0)$ order does not collapse abruptly, but instead gradually evolves into a short-range ordered glassy sate through a crossover transition. The AFM transition observed in the resistivity and specific heat data of the $x = 0.05$ sample of the $Fe_{1.1}$ series (see Figs. 7(a) and 8(a)) is indeed very broad, consistent with a crossover transition. In this case, we could imagine that the incoherent magnetic scattering would not have substantial increase for $x > 0.1$. This might be the reason why the $\gamma$ enhancement near $x \sim 0.1$ is much smaller in the $Fe_{1.1}$ system than in the $Fe_{1.02}$ system (Fig. 1(b)) and again reflects the dependence of band renormalization on the magnetic scattering.

**B.2. Hall effect of $Fe_{1.1}(Te_{1-x}Se_x)$**

The Hall coefficient $R_H$ data we collected for the $Fe_{1.1}$ samples, as shown in Fig. 9, provides further support for our interpretation for the specific heat data. Compared to the $R_H$ data of the Fe$_{1.02}$ system (Fig. 4), the Fe$_{1.1}$ system exhibits the following distinct features in $R_H$: a) For $x < 0.1$ region, $R_H$ remains positive down to 2 K though it shows a drop across the AFM transition, in contrast with the sign reversal of $R_H$ across the AFM transition seen in the Fe$_{1.02}$ system; b) For $x > 0.1$ region, $R_H$ shows monotonic increase with decreasing temperature and remains positive even for $x > 0.3$ (see Fig. 9(b)), in sharply contrast with that seen in the Fe$_{1.02}$ system where $R_H$ displays a non-monotonic temperature dependence, with a crossover from positive to negative at low temperature for $x > 0.38$. We have added the normalized $R_H$ data at 20 K of the $y \sim 0.1$ system to Fig. 1(c) for comparison with the Fe$_{1.02}$ system, which clearly shows that the increase of interstitial Fe leads low-temperature $R_H$ to change from negative to positive in both the $x < 0.1$ AFM region and the $x > 0.38$ region. This observation agrees well with our argument proposed above: the magnetic scattering by $(\pi,0)$ magnetic fluctuations reduces electron mobility more effectively. Since strong $(\pi,0)$ magnetic correlation extends to the high Se concentration region for the Fe$_{1.1}$ system as discussed above, the magnetic scattering remains strong even for $x > 0.38$, thus resulting in low electron mobility. In this case, the transport may be dominated by holes, which leads to positive $R_H$. Positive $R_H$ below $T_N$ for $x < 0.1$ should have a similar mechanism, since rich interstitial Fe is expected to enhance incoherent magnetic scattering though the system is in an AFM ordered state as mentioned above. Although we do not know how strongly the magnetic scattering affects hole mobility, our results, at minimum, suggest that the magnetic scattering causes greater reduction in mobility for electrons than for holes.

### B.3. Magnetoresistance of Fe$_{1.1}$(Te$_{1-x}$Se$_x$)

We have shown above that the magnetic scattering leads to negative MR in the $0.1 < x < 0.3$ region of the $Fe_{1.02}$ series. Given that rich interstitial Fe enhances the $(\pi,0)$ magnetic correlation and that strong $(\pi,0)$ magnetic fluctuations extend to the $x > 0.3$ region for the $Fe_{1.1}$ series, the negative MR arising from the magnetic scattering can be expected even for $x > 0.3$ in the $Fe_{1.1}$ system. This is exactly what we observed in our experiments. As shown in Fig. 10, the MR is positive only in the $x < 0.1$ AFM region but becomes negative for all of the samples with $x > 0.1$ (see the inset to Fig. 10). Furthermore, we note that the magnitude of MR hardly changes with Se concentration for $x > 0.1$; this is consistent with the above argument that the $(\pi,0)$ magnetic correlation does not weaken noticeably with the increase of Se concentration in the $Fe_{1.1}$ system.

Another distinct feature seen in the MR data of Fig. 10 is that the positive MR of parent compound $Fe_{1.1}Te$ linearly depends on magnetic field for $B > 2$ T and its magnitude is much larger than that of the rest of the samples. Such a linear field dependence of MR contrasts with the small $B^2$-like MR observed in $Fe_{1.02}Te$ (Fig. 6(a)), but looks similar to what has been observed in iron pnictides [55-57], where the linear field dependence of MR is attributed to Dirac cone states [58]. This observation implies that the increase of interstitial Fe leads to significant change of band structure. Our observation of the increase of $\gamma$ from ~ 30 mJ/mol K$^2$ for $Fe_{1.02}Te$ to ~ 40 mJ/mol K$^2$ for $Fe_{1.1}Te$ (see Fig. 1(b)) suggests that this is the most possible scenario [58]. However, it is not clear whether the linear field dependence of MR in $Fe_{1.1}Te$ corresponds to Dirac cone states as suggested in iron pnictides. This question certainly deserves further investigation.

**IV. Conclusion**

In summary, we performed systematic measurements on specific heat, $R_H$, and MR for $Fe_{1+y}(Te_{1-x}Se_x)$ ($0 \leq x \leq 0.5$; $y = 0.02$ and $0.1$). In the $y = 0.02$ system, we observed that $\gamma$, $R_H$ and MR all show remarkable variations with Se concentration and depend on the strength of $(\pi,0)$ magnetic correlation. The $0.1 < x < 0.3$ region, where the $(\pi,0)$ magnetic correlation is strong and bulk superconductivity is suppressed, is characterized by large $\gamma$ ($\sim$ 55-65 mJ/mol K$^2$), positive $R_H$ and negative MR at low temperatures, while, the $x = 0.4$-$0.5$ region, where the $(\pi,0)$ magnetic correlation is significantly suppressed, features relatively small $\gamma$ ($\sim$ 26mJ/mol K$^2$), negative $R_H$ and positive MR at low temperatures. Our analyses show that the most possible mechanism for such coupling between electronic properties and the $(\pi,0)$ magnetic correlation is that the $(\pi,0)$ magnetic fluctuations give rise to significant incoherent magnetic scattering in the $0.1 < x < 0.3$ range, which induces incoherence of electronic states and consequently causes superconducting pair breaking and charge carrier localization. In the $x = 0.4$-$0.5$ region, the magnetic scattering by the $(\pi,0)$ magnetic fluctuations, while it still exists, is significantly weakened, which leads electron mobility to increase strikingly, thus resulting in bulk superconducting pairing and normal state metallic transport. The enhanced $\gamma$ in the $0.1 < x < 0.3$ region implies the relevance of band renormalization with the magnetic scattering. The dependence of electronic properties on the $(\pi,0)$ magnetic correlation is also verified in the system with rich interstitial Fe ($y = 0.1$). Since interstitial Fe increases the $(\pi,0)$ magnetic correlation length, the increase of interstitial Fe strengthens the $(\pi,0)$ short-range order so that the increase of Se content does not effectively weaken the $(\pi,0)$ short-range order even for $x > 0.3$ in the $y = 0.1$ system as it does in the $y = 0.02$ system. In this case, the weakly localized state caused by magnetic scattering extends to high Se concentrations. The weak variations of $\gamma$, $R_H$, and MR in the $y = 0.1$ system all can be well understood in light of the less striking evolution of magnetic scattering. In addition, we find that the magnetic scattering can be suppressed by magnetic field, which can lead to superconductivity enhancement in the $x = 0.3$-$0.4$ samples of the $y = 0.02$ system.

**Acknowledgments**

The work is supported by the NSF under grant DMR-1205469 and the LA-SiGMA program under award #EPS-1003897. The authors are grateful to W. Ku for informative discussions and D. Fobes for technical support.

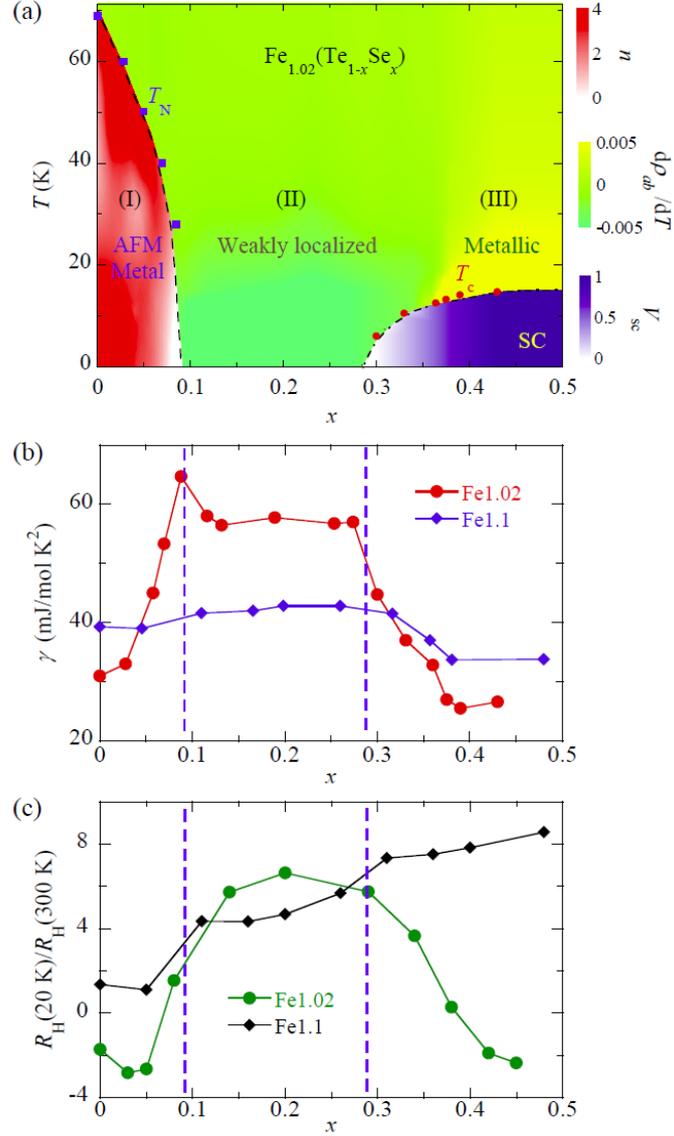

**Figure 1**. (a) The magnetic and electronic phase diagram of $Fe_{1.02}(Te_{1-x}Se_x)$ ($0 \leq x \leq 0.5$). $T_N$ and $T_c$ represent the Néel temperature and onset superconducting transition temperature probed by specific heat (see Fig. 2), respectively, which agree well with the $T_N$ (dashed line) and $T_c$ (dash-dotted line) phase boundaries previously determined using neutron scattering and magnetic susceptibility measurements [10]. The contour plots illustrate the variation of the exponent $n$ of the temperature–dependent term of in-plane-

resistivity $\rho_{ab}$ ($\propto T^n$) for the AFM phase, the derivative of $\rho_{ab}$ for Region II and the normal state of Region III, and superconducting volume fraction $V_{SC}$ estimated from specific heat for the superconducting phase (see text), respectively. (b) and (c): Sommerfeld coefficient $\gamma$ and normalized Hall coefficient $R_H$ as a function of Se concentration $x$ at 20 K for $Fe_{1.02}(Te_{1-x}Se_x)$ and $Fe_{1.1}(Te_{1-x}Se_x)$ samples (denoted as $Fe_{1.02}$ and $Fe_{1.1}$ in the figure). For $Fe_{1.02}(Te_{1-x}Se_x)$ samples, $\gamma$ in Regions I and II and $R_H$ in Region I are taken from our previous work [10].

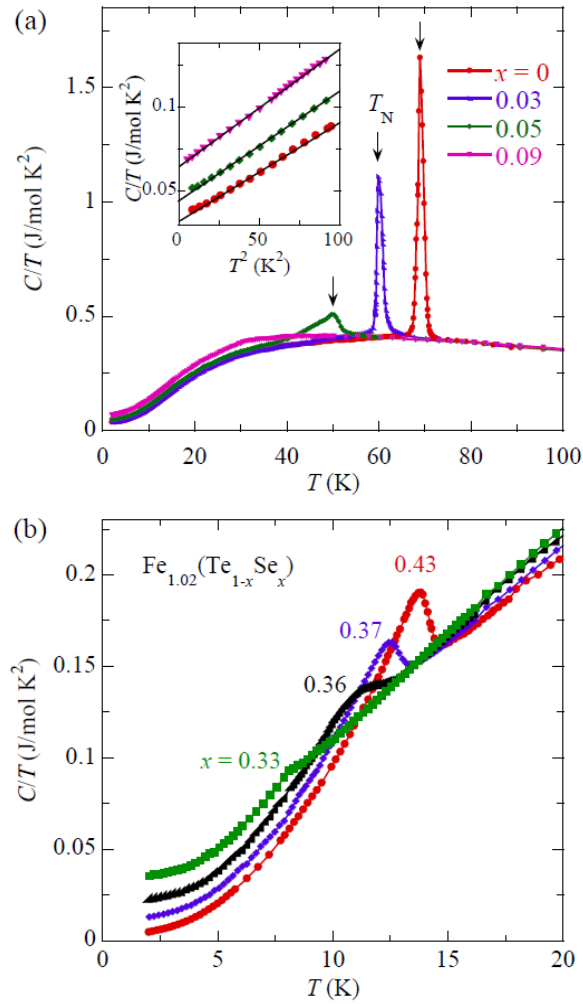

**Figure 2**. Specific heat divided by temperature $C/T$ as a function of temperature for the Fe$_{1.02}$(Te$_{1-x}$Se$_x$) samples in (a) $0 \leq x < 0.1$ (Region I) and (b) $0.3 < x < 0.5$ (Region III). The inset in (a) shows $C/T$ vs. $T^2$, with the solid lines representing linear fits.

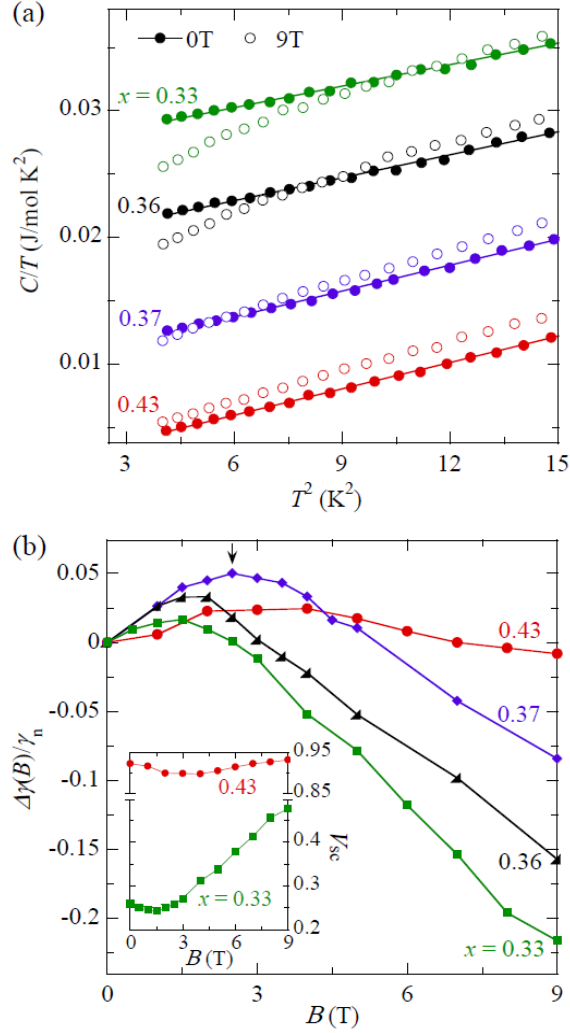

**Figure 3**. (a): $C/T$ vs. $T^2$ at $B = 0$ T and 9 T ($B//c$-axis) for $Fe_{1.02}(Te_{1-x}Se_x)$ samples with x = 0.33, 0.36, 0.37 and 0.43. The solid lines represent the linear fits for the zero field data. No sign of magnetic or Schottky anomaly is observed down to 2 K. (b): The field-induced change in electronic specific heat $\Delta\gamma/\gamma$ ($\Delta\gamma=\gamma(B)-\gamma(0)$) at the zero temperature limit as a function of field. The arrow indicates the field where $\Delta\gamma$ reaches a maximum; Inset: the superconducting volume fraction $V_{SC}$ as a function of field for $x = 0.33$ and 0.43 samples.

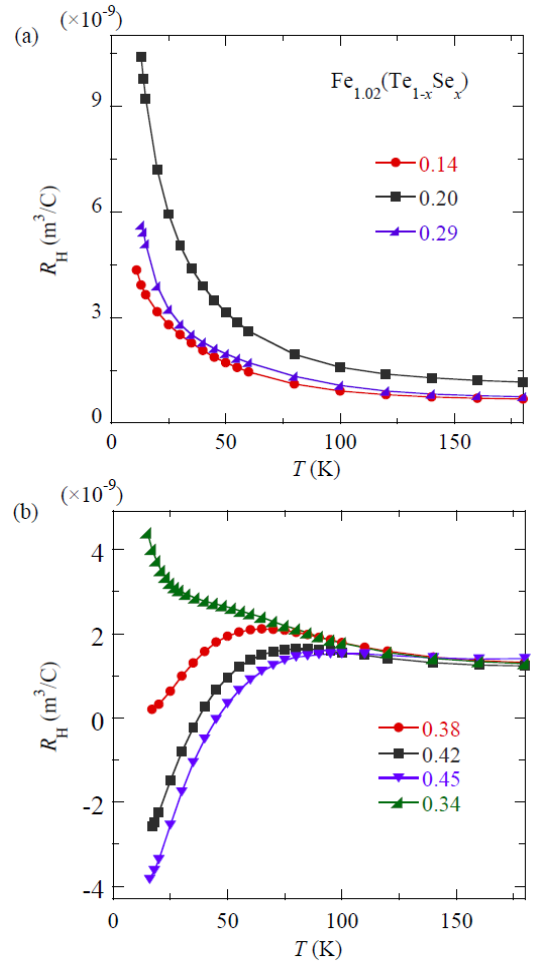

**Figure 4.** Temperature dependence of Hall coefficient $R_H$ for $Fe_{1.02}(Te_{1-x}Se_x)$ samples with (a) $0.1 < x < 0.3$ (Region II in Fig. 1(a)) and (b) $0.3 < x < 0.5$ (Region III).

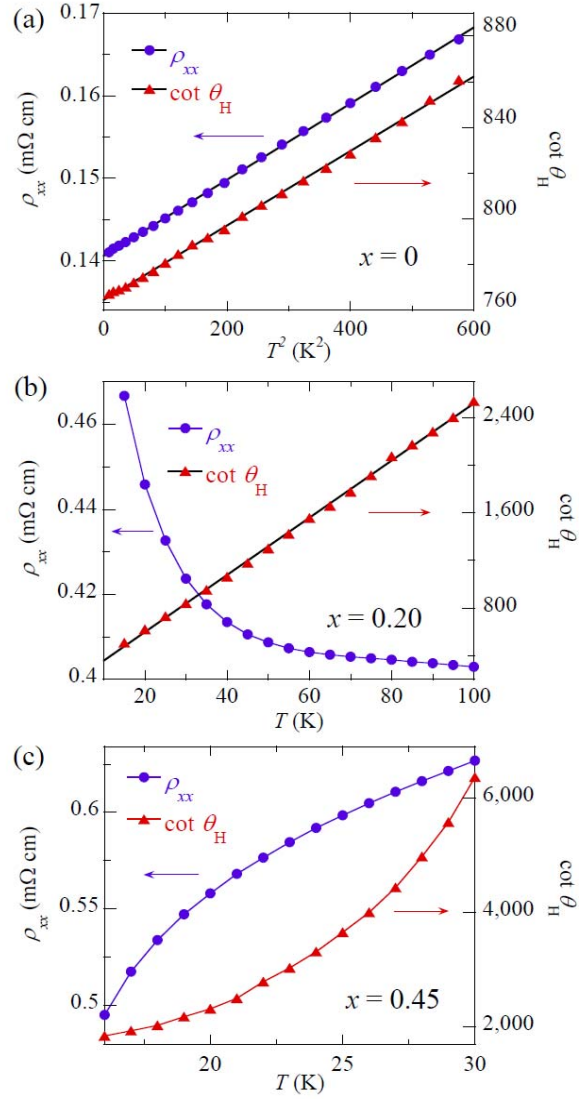

**Figure 5.** Temperature dependence of in-plane resistivity $\rho_{xx}$ and Hall angle $\cot\theta_H$ for typical samples with $x = 0$ (c), 0.20 (d), and 0.45 (e).

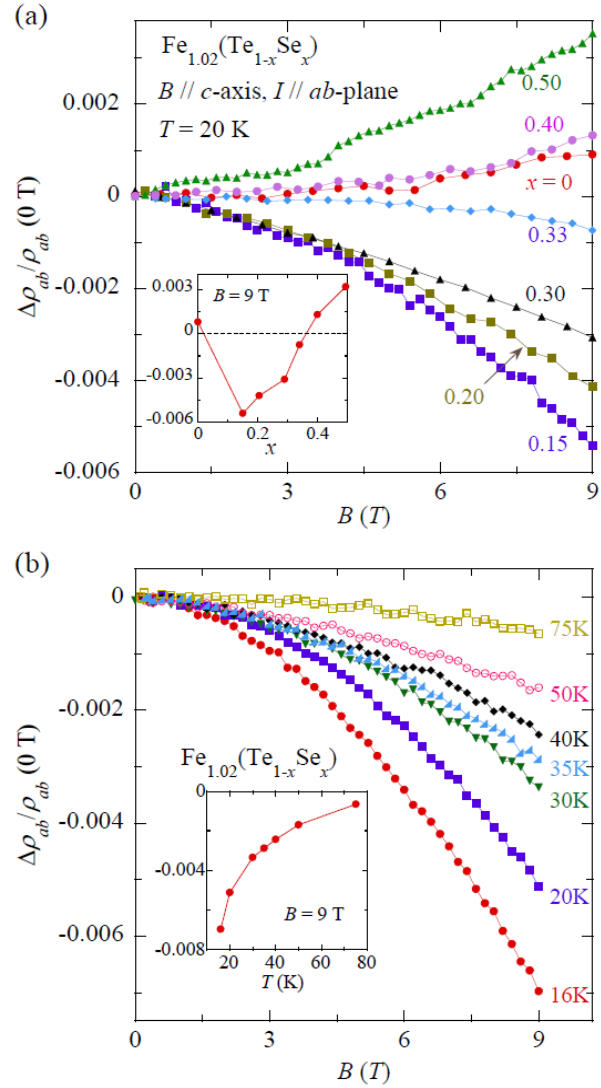

**Figure 6.** (a) Field dependence of MR, $\Delta\rho_{ab}/\rho_{ab}(0T) = [\rho_{ab}(B)-\rho_{ab}(0T)]/\rho_{ab}(0T)$ at 20 K for $Fe_{1.02}(Te_{1-x}Se_x)$ samples with various Se concentrations; Inset, $\Delta\rho_{ab}/\rho_{ab}(0T)$ at 20 K vs. Se content $x$ under the applied magnetic field of 9 T. (b) Field dependence of MR at various temperatures for $Fe_{1.02}(Te_{0.85}Se_{0.15})$ sample; Inset, $\Delta\rho_{ab}/\rho_{ab}(0T)$ vs. temperature under the applied magnetic field of 9 T for $Fe_{1.02}(Te_{0.85}Se_{0.15})$ sample.

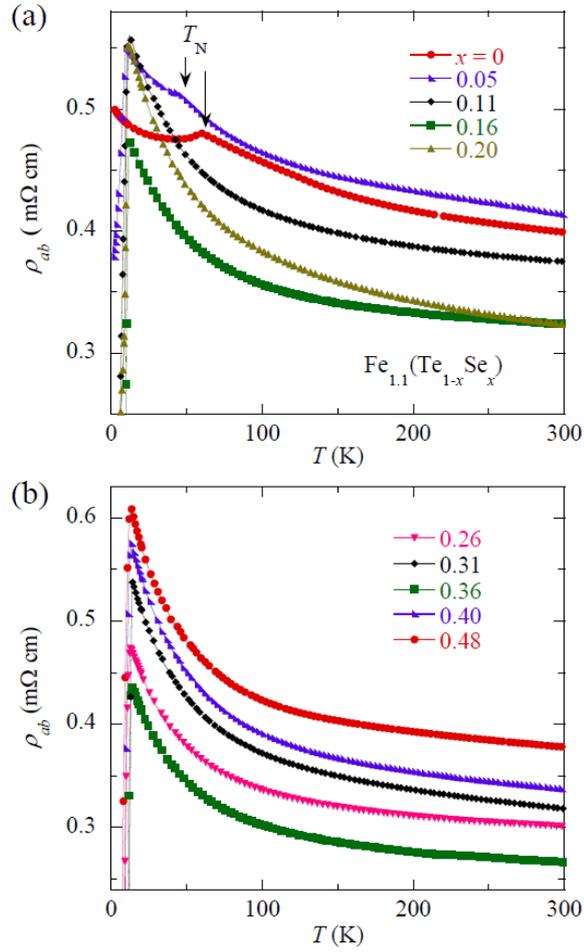

**Figure 7.** Temperature dependence of in-plane resistivity $\rho_{ab}$ for $Fe_{1.1}(Te_{1-x}Se_x)$ samples with (a) $0 \leq x \leq 0.2$ and (b) $0.2 < x < 0.5$.

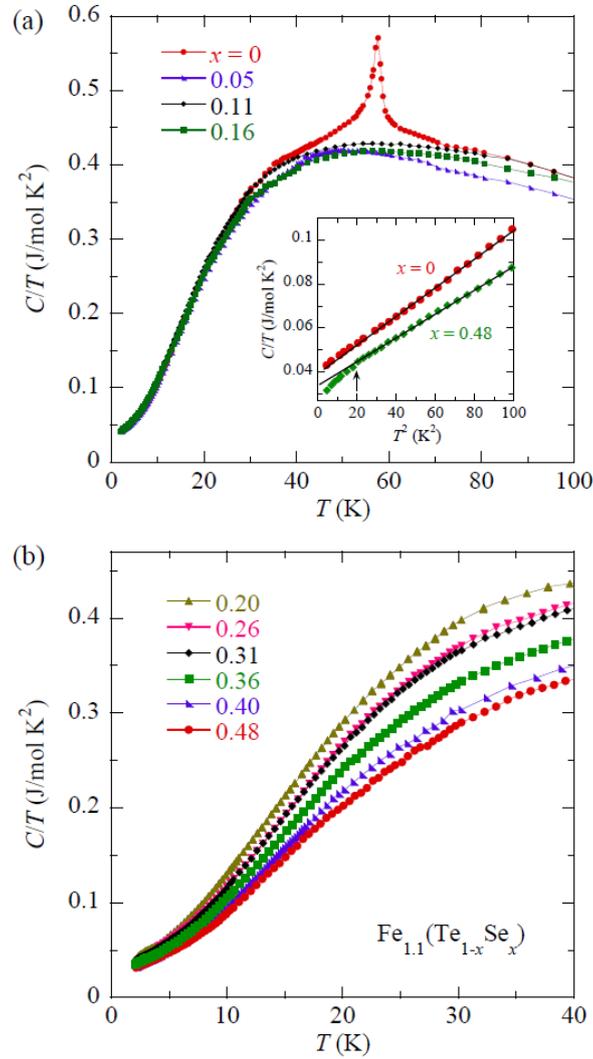

**Figure 8.** Specific heat divided by temperature $C/T$ as a function of temperature for $Fe_{1.1}(Te_{1-x}Se_x)$ samples with (a) $0 \leq x < 0.2$ and (b) $0.2 \leq x < 0.5$. The inset in (a) shows the linear fit of $C/T$ vs. $T^2$ for $x = 0$ and $0.48$ samples; the upward arrow indicates a deviation from linearity for the $x = 0.48$ sample due to non-bulk superconductivity.

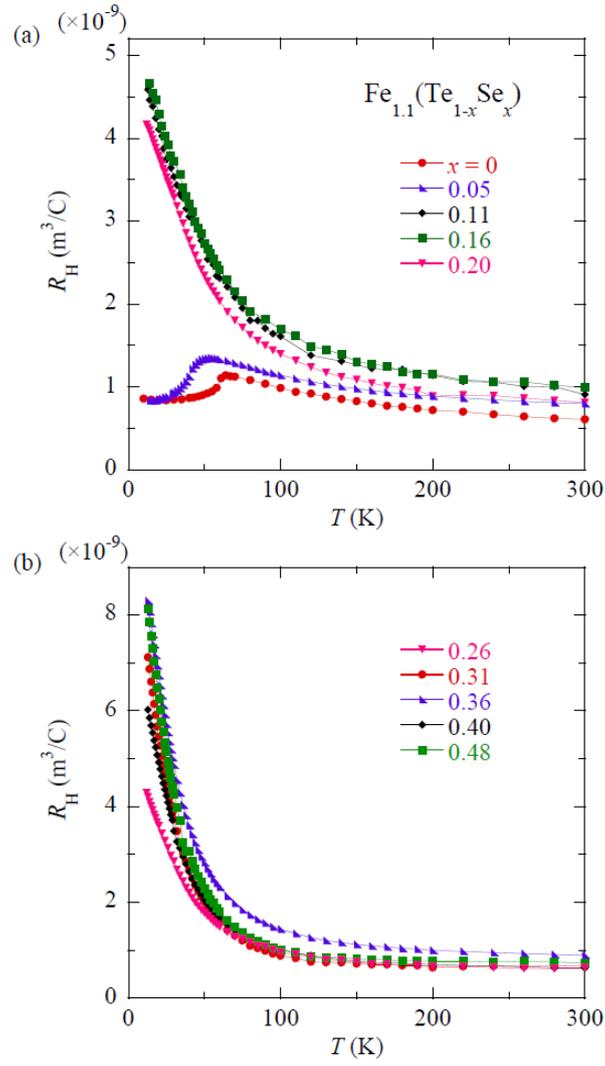

**Figure 9.** Temperature dependence of Hall coefficient $R_H$ for $Fe_{1.1}(Te_{1-x}Se_x)$ samples with (a) $0 \leq x \leq 0.2$ and (b) $0.2 < x < 0.5$.

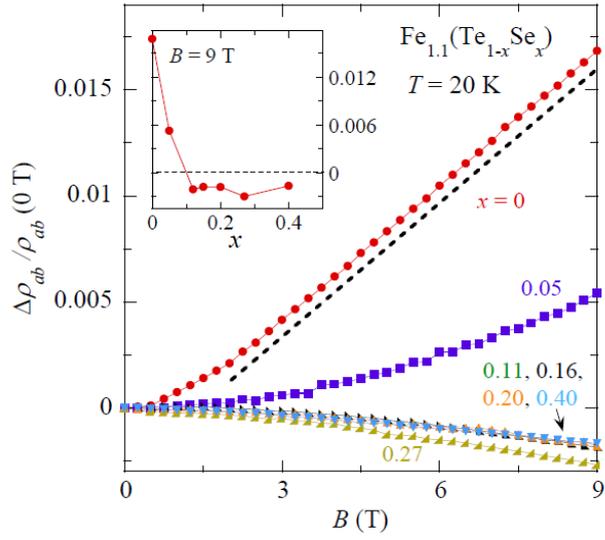

**Figure 10.** (a) Field dependence of MR, $\Delta\rho_{ab}/\rho_{ab}(0T) = [\rho_{ab}(B)-\rho_{ab}(0T)]/\rho_{ab}(0T)$ at 20 K for $Fe_{1.1}(Te_{1-x}Se_x)$ samples with various Se concentrations. The dashed marks the linear field dependence of MR for $x = 0$ sample. Inset, $\Delta\rho_{ab}/\rho_{ab}(0T)$ at 20 K *vs*. Se content $x$ under the applied magnetic field of 9 T.